\documentclass [prl,amsmath,showpacs,twocolumn,preprintnumbers,superscriptaddress]{revtex4-1}

\usepackage{mathptmx}
\usepackage{microtype}
\usepackage[latin1]{inputenc}
\usepackage[spanish]{babel}
\usepackage{graphicx}
\usepackage{dcolumn}
\usepackage{amsmath}
\usepackage{mathtools}
\usepackage{bm}
\usepackage{amssymb}

\begin{document}

\title{Termodinámica del universo temprano desde una cosmología newtoniana}

\author{W. A. Rojas C.}
\email{warojasc@unal.edu.co}
\affiliation{Universidad Nacional de Colombia}

\begin{abstract}
We study a model of fluent material universe from a modified Newtonian cosmology proposed by D'Inverno \cite {D'Inverno} and Tawfik \cite{Tawfik}. From this perspective, it shows a list of space-time curvature $ k $ with energy $ E $, density $ \rho $, specific heat $ C_{p} $ and the fluid temperature gradient considered $ \Delta T $ corresponding to standard thermodynamic variables. We determined the coupling constants accompanying the matter density for different values we can take $ k $.
\end {abstract}


\maketitle
\section{Introducción}
Consideremos una cosmología newtoniana sin constante cosmológica, tal como lo plantea Inverno\cite{D'Inverno},  asumiendo que el universo se compone de una cierta cantidad materia en el sentido clásico, tal que la posición  y velocidad de  todas las partículas se pueden medir desde cierto punto $O$. Ademas de imponer que el Principio Cosmológico sea válido en este contexto. Tawfik muestra que al estudiar este tipo de cosmología y teniendo en cuenta los efectos viscosos de un modelo de universo se puede obtener resultados interesantes que son concordantes que la Cosmología FRW \cite{Tawfik}. De acuerdo a lo anterior se plantea la necesidad de investigar los efectos que experimentan los parametros  termodinámicos  como lo son la densidad $\rho$ y el calor específico $C_{p}$ en función de de parametros cosmológicos tales como el factor de escala $a(t)$ y la constante de Hubble $H_{0}$.

\section{Un modelo de cosmología newtoniana}
De acuerdo a lo anterior se tiene que  la energía total de una partícula de masa $m$ que se halla a una distancia $a$ de un punto arbitrario $O$, que se puede mover con cierta velocidad radial $\dot{a}$ y  está afectada por el potencial gravitacional $V(a)=-\frac{GMm}{a}$; con $G$ siendo la constante de gravitación universal y $M$ la masa del fluido en el cual se mueve la partícula e igual a $M=\frac{4}{3}\pi a^{3}\rho$. Por lo que el fluido experimenta una ganancia de energía térmica debído al movimiento de la partícula dentro de el \cite{D'Inverno, Tawfik} 
\begin{equation}
E=\frac{1}{2}m\dot{a}^{2}-\frac{GMm}{a}-C_{p}M\Delta T.
\end{equation}
Con $C_{p}$ siendo el calor específico del fluido y $\Delta T$ el gradiente de temperatura al cual se somete la partícula. Asi (1) se puede reescribir como
\begin{equation}
\dot{a}^{2}-\frac{2E}{m}-\frac{8}{3}\frac{a^{3}\rho C_{p}\Delta T}{m}=\frac{8}{3}\pi G a^{2}\rho.
\end{equation}
Comparando con la primera de las ecuaciones de Friedmann derivadas de la Relatividad General sin constante cosmológica para un modelo de gas ideal
\begin{equation}
\dot{a}^{2}+k=\frac{8}{3}\pi G a^{2}\rho,
\end{equation}
de acuerdo a lo anterior se logra vincular la curvatura espacio-temporal $k$ con ciertas variables termodinámicas tal como el calor específico $C_{p}$, la densidad $\rho$ y  el gradiente de temperatura $\Delta T$
\begin{equation}
k=-\frac{2E}{m}-\frac{8}{3}\frac{a^{3}\rho C_{p}\Delta T}{m}.
\end{equation}
Con los posibles valores que puede tomar $k$ $[+1,0,-1]$ se tendrán los  diferentes escenarios de curvatura del espacio-tiempo:
\begin{enumerate}
	\item En el caso de $k=+1$ con lo que (4) se puede determinar el valor de $\rho$ en terminos de los demas parámetros
\begin{equation}
\rho=-\frac{3}{8} \frac{(m+2E)}{C_{p}\Delta T} \frac{1}{a^{3}}
\end{equation}
\item En el caso de $k=-1$ se  tiene
\begin{equation}
\rho=\frac{3}{8} \frac{(m-2E)}{C_{p}\Delta T} \frac{1}{a^{3}}
\end{equation}
 \item En el caso de $k=0$ se  tiene
\begin{equation}
\rho=\frac{-E}{4C_{p}\Delta T} \frac{1}{a^{3}},
\end{equation}
\end{enumerate}
En todos los  casos considerados se muestra que $\rho \propto \frac{1}{a^{3}}$ lo cual indica un modelo de universo dominado por materia, que era lo que se pretendia desde el principio. Consideraciones para (5), (6) y (7) la densidad debe ser simpre mayor que cero. La masa de igual forma es mayor  que cero; no se tienen en cuenta formas extrañas de materia (masa o energía oscura).  $\Delta T<0$ pues $\Delta T= T_{f}-T_{0}$ pues $T_{0}>T_{f}$ pues el universo comenzó en un punto de muy elevada densidad y temperatura. El calor específico del sistema se comporta clásicamente $C_{p}>0$.

\section{Cosmología FRW}
Para un espacio-tiempo tipo FRW \cite{McMahon} se tiene que 
	\[\nabla_{\mu}T^{\mu \nu}=0,
\]
donde $T^{\mu \nu}$ es el tensor momentum-energía para un gas ideal, y el hecho que su derivada covariante sea igual a cero esta asociada con la primera ley de termodinámica, de lo que se desprende
\begin{equation}
\frac{\partial \rho}{\partial t}=-\frac{3 \dot{a}}{a}(\rho+ P),
\end{equation}
con $P$ siendo la presión ejercida por el gas ideal. Si elegimos $P=0$, (8) se reduce a 
\begin{equation}
\rho=\frac{cte}{a^{3}}.
\end{equation}
De la comparación de (9) con (5), (6) y (7) se puede determinar los posibles valores que puede tomar tal constante para los casos de $k$ considerados
\begin{enumerate}
	\item Con $k=1$ 
		\[cte=-\left(\frac{3}{8}\right)\frac{(m+2E)}{C_{p}\Delta T}.
\]
	\item Con $k=-1$ 
		\[cte=\left(\frac{3}{8}\right)\frac{(m-2E)}{C_{p}\Delta T}.
\]
	\item Con $k=0$ 
		\[cte=-\frac{2E}{4C_{p}\Delta T}.
\]
\end{enumerate}
Donde hemos asumido una ecuación de estado de la forma
	\[\rho \propto a^{-3(w+1)},
\]
que para el caso que estamos estudiando $w=0$; por lo que $\rho \propto a^{-3}$\cite{Carroll}. Si establecemos la hipótesis que la densidad del fluido en nuestro modelo sea igual a la densidad critica; con la densidad crítica definida como 
	\[\rho_{c}=\frac{3H^{2}_{0}}{8 \pi G}
\]
donde $H_{0}$ corresponde a la constante de Hubble para la época actual. Por lo que podemos hallar el calor específico del fluido en consideración de acuerdo a los valores que toma $k$
\begin{enumerate}
	\item Para $k=1$
	\begin{equation}
	C_{p1}=-\frac{\pi G(m+2E)}{\Delta TH^{2}_{0}}\left( \frac{1}{a^{3}}\right).
	\end{equation}
	\item Para $k=-1$
	\begin{equation}
	C_{p2}=\frac{\pi G(m-2E)}{\Delta TH^{2}_{0}} \left( \frac{1}{a^{3}}\right).
	\end{equation}
	\item Para $k=0$
	\begin{equation}
	C_{p3}=-\frac{2\pi GE}{3\Delta TH^{2}_{0}} \left( \frac{1}{a^{3}}\right).
	\end{equation}
\end{enumerate}
\begin{figure}[h]
	\centering
		\includegraphics[width=0.4\textwidth]{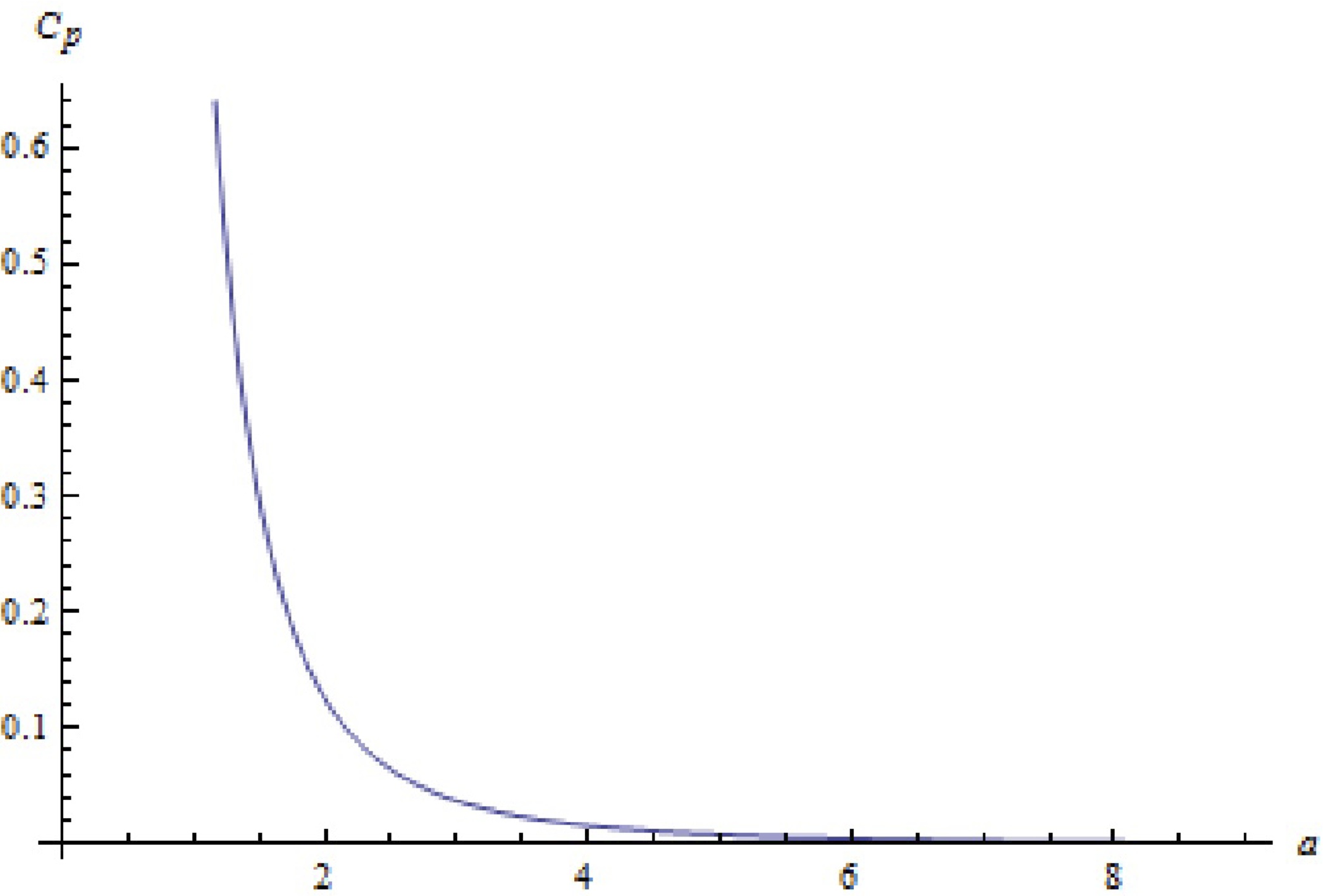}
		\caption{Comportamiento del calor específico $C_{p}$ en función del factor de escala $a(t)$}
	\label{fig:triangulo}
\end{figure}
En los casos analizados se observa que $C_{p}\propto \frac{1}{a^{3}}$, lo cual significa que el calor específico para universo con dominio de materia decrece al expandirse el universo; tal como se aprecia en la Figura 1. La relación existente entre $C_{p1}/C_{p2}=-1$ lo que permite afirmar que la geometría espacio-temporal afecta el calor específico del fluido en consideración.

\section{Conclusiones}
El uso de una cosmología newtoniana ha mostrado resultados consistentes con la cosmología FRW. Así, la curvatura espacio-tiempo está relaciona con la energía $E$, la densidad $\rho$, el calor específico $C_{p}$ y el gradiente de temperatura del fluido considerado $\Delta T$ que corresponden a variables de la  termodinámica estandar. Cuando se consideró las constantes de acople que acompañan la densidad descritas en (5), (6) y (7) para los  diferentes valores, que puede tomar $k$ son consistentes con las soluciones estandard de la cosmología. 

Un hecho relavante es  que el calor específico esta dado como $C_{p}\propto \frac{1}{H^{2}_{0}a^{3}}$ en los casos considerados; este decrece al expandirse y enfriarse el universo.

\section*{References}

\providecommand{\newblock}{}


\end{document}